\documentclass[10pt,a4paper,english]{revtex4}
\usepackage[T1]{fontenc}
\usepackage[latin9]{inputenc}
\usepackage{amsbsy}
\usepackage{graphicx}
\usepackage{esint}

\makeatletter

\providecommand{\tabularnewline}{\\}

\@ifundefined{textcolor}{}
{%
 \definecolor{BLACK}{gray}{0}
 \definecolor{WHITE}{gray}{1}
 \definecolor{RED}{rgb}{1,0,0}
 \definecolor{GREEN}{rgb}{0,1,0}
 \definecolor{BLUE}{rgb}{0,0,1}
 \definecolor{CYAN}{cmyk}{1,0,0,0}
 \definecolor{MAGENTA}{cmyk}{0,1,0,0}
 \definecolor{YELLOW}{cmyk}{0,0,1,0}
}

\usepackage{babel}

\makeatother

\usepackage{babel}

\makeatother

\usepackage{babel}

\makeatother

\usepackage{babel}

\makeatother

\usepackage{babel}

\makeatother

\usepackage{babel}
\begin{document}

\title{Detecting stable massive neutral particles through particle lensing}

\author{Luca Amendola$^{1}$ \& Valeria Pettorino$^{2,3}$}

\affiliation{$^{1}$ Institut fuer Theoretische Physik, Universitaet Heidelberg,
Philosophenweg 16, D-69120 Heidelberg, Germany. \\
 $^{2}$ D\'epartement de Physique Th\'eorique, Universit\'e de Gen\`eve,
24 quai Ernest Ansermet, 1211, Gen\`eve 4, Switzerland\\
 $^{3}$ SISSA, Via Bonomea 265, 34136 Trieste, Italy, \\
 }
 
\begin{abstract}
Stable massive neutral particles emitted by astrophysical sources
undergo deflection under the gravitational potential of our own galaxy.
The deflection angle depends on the particle velocity and therefore
non-relativistic particles will be deflected more than relativistic
ones. If these particles can be detected through neutrino telescopes,
cosmic ray detectors or directional dark matter detectors, their arrival
directions would appear aligned on the sky along the source-lens direction.
On top of this deflection, the arrival direction of non-relativistic
particles is displaced with respect to the relativistic counterpart
also due to the relative motion of the source with respect to the
observer; this induces an alignment of detections along the sky projection
of the source trajectory. The final alignment will be given by a combination
of the directions induced by lensing and source proper motion. We
derive the deflection-velocity relation for the Milky Way halo and
suggest that searching for alignments on detection maps of particle
telescopes could be a way to find new particles or new astrophysical
phenomena. 
\end{abstract}
\maketitle

\section{Introduction}

\begin{figure}
\includegraphics[bb=0bp 260bp 580bp 790bp,clip,scale=0.6]{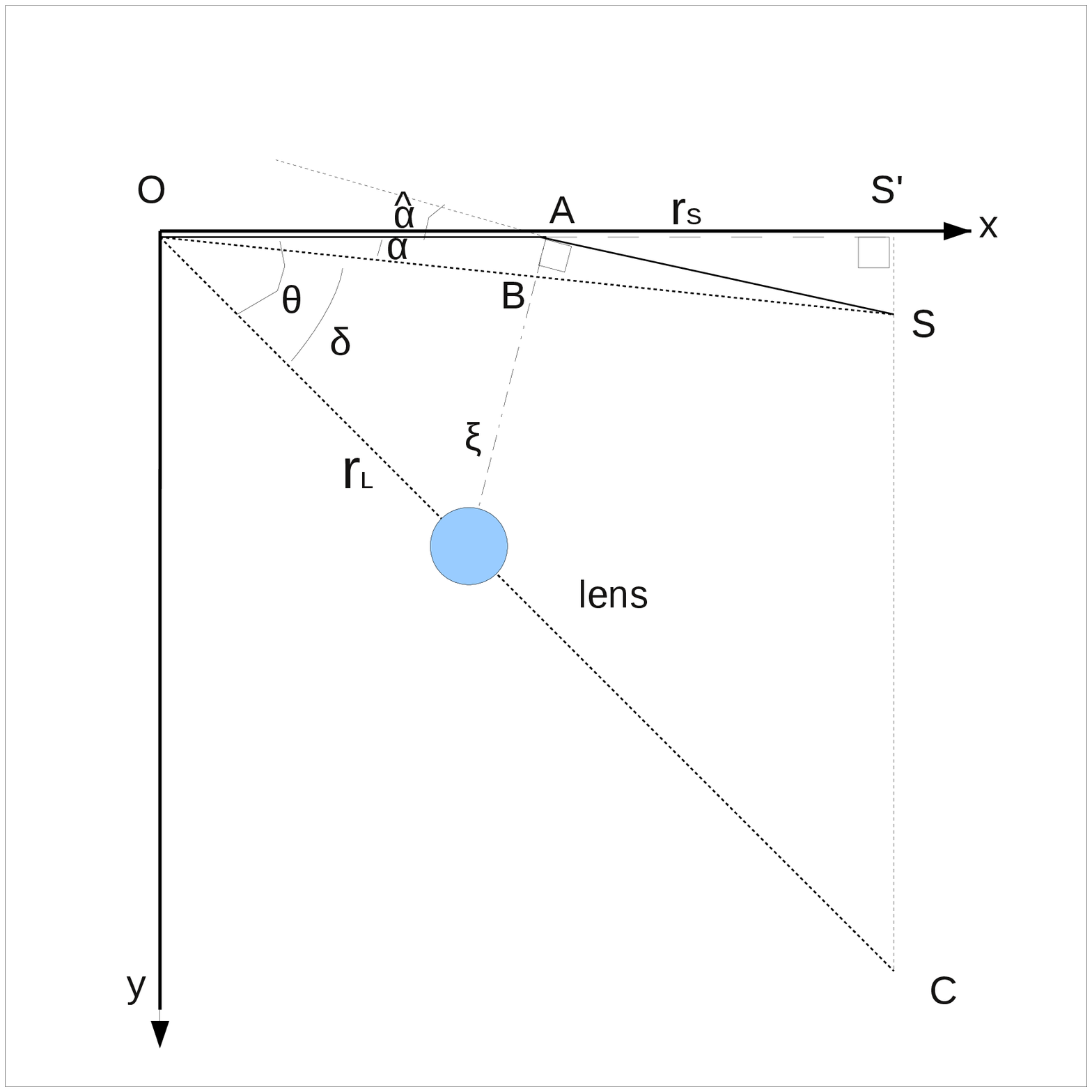}

\caption{\label{fig:Angles-1}The geometric configuration of the halo lens
system. The observer is in O, the source in S, its image in S', and
the lens is the filled circle. Point C is added to ease the discussion.
In the following we always assume $\alpha,\hat{\alpha}$ to be small.
The angles $\theta,\delta$ are also small in Sect. \ref{sec:Particle-lensing}
while are left free in Sect. \ref{sec:Lensing-in-a}.}
\end{figure}

As it is well-known (see e.g. \cite{1992grle.book.....S}), a light
ray trajectory is deviated by a compact mass $M$ by an amount 
\begin{equation}
\hat{\alpha}=\frac{2R_{S}}{\xi}\label{eq:rlens}Ä
\end{equation}
 where $R_{S}=2GM/c^{2}$ is the Schwarzschild radius of the deflector
(or ``lens'') and $\xi$ is the impact parameter, i.e. the shortest
distance of the particle to the deflector along the imperturbed path.
The angle $\hat{\alpha}$, denoted lensing angle, defined as in Fig.
(\ref{fig:Angles-1}), lies in the plane observer-lens-source. This
equation holds true for small lensing angles and for a deflector size
much smaller than the source distance.

The analogous deflection for a non-relativistic (NR) particle is easy
to derive in the Newtonian limit. If a particle is emitted by an astrophysical
source with a non-relativistic velocity $v$ (in units of $c$), its
trajectory will be deflected by any mass close to its trajectory by
an angular amount equal to (e.g. \cite{1992grle.book.....S} p. 2)
\begin{equation}
\hat{\alpha}=\frac{R_{S}}{v^{2}\xi}\label{eq:nrlens}
\end{equation}
 again in the limit of small deflection angles and deflector size.
Crucially, this deflection is amplified with respect to the relativistic
one by the $v^{-2}$ factor. Because of the strict analogy, we refer
to this deflection as particle  lensing, although in neither case
there is any focusing of images.

\begin{figure}
\includegraphics[bb=0bp 320bp 595bp 842bp,clip,scale=0.5]{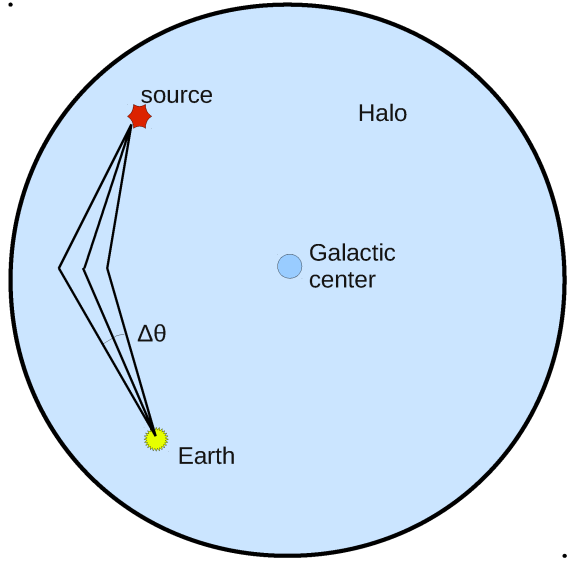}\caption{\label{fig:Schematic-representation-of}Schematic representation of
the deflection due to the galactic halo.}
\end{figure}

\begin{figure}
\includegraphics[bb=0bp 330bp 580bp 890bp,clip,scale=0.6]{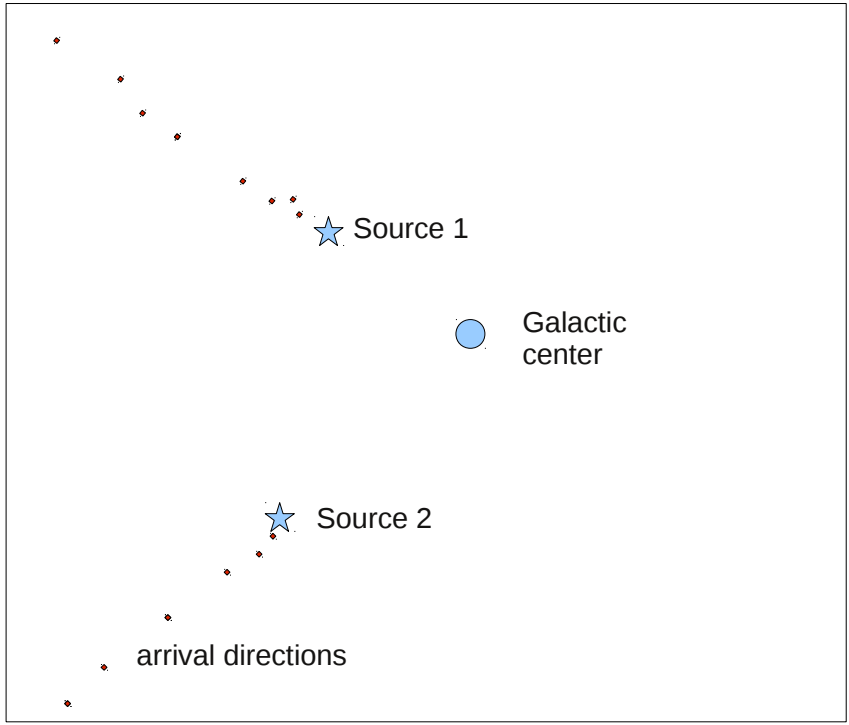}

\caption{\label{fig:Angles-2}A depiction of the detection map with non-relativistic
lensing for a source with negligible proper motion, purely for illustrative
purposes. The little red dots represent the arrival directions of
massive particles emitted by source 1 and 2.}
\end{figure}

If the particle is neutral, its trajectory will not be further deflected
by the galactic magnetic field and by estimating $\hat{\alpha},\xi$
and $R_{s}$ we can infer the particle velocity and, given the kinetic
energy, its mass. If particles of the same kind are emitted by an
astrophysical source with a range of velocities, they will all travel
on the same plane observer-lens-source and reach the observer with
different angles $\hat{\alpha}(v)$. The observer will therefore detect
particles with different energies aligned on the sky in the direction
source-lens (see Figs. \ref{fig:Schematic-representation-of},\ref{fig:Angles-2}).
In this paper we suggest that this peculiar alignment of detections
in particle telescopes, as e.g. neutrino telescopes (e.g. \cite{2011NIMPA.630..125P})
cosmic ray detectors (e.g. \cite{2011arXiv1107.4809T}) or future
directional dark matter detectors (e.g. \cite{2010IJMPA..25....1A,2009MPLA...24.1793S}),
could be used as a test to reveal the existence of stable massive
neutral particles (SMaNPs) emitted by astrophysical sources inside
or outside our Galaxy. Let us remark from the outset that although
the method we propose is extremely simple, it is not easy to identify
possible targets for its application within the current frontiers
of particle physics and astrophysics and it is therefore to be meant
as a search for unexpected phenomena. For this reason, we will not
investigate in any detail how and where such neutral particles could
be produced and accelerated.

Since dark matter particles are also expected to be SMaNPs, it is
clear that a novel way of detecting them would be potentially very
interesting. However, the particle lensing method applies only to
particles emitted by astrophysical bodies and clearly cannot be employed
to detect halo dark matter particles. Possible candidates are massive
sterile neutrinos (e.g. \cite{2009PhLB..670..281F}) or exotic components
of cosmic rays (e.g. heavy axion-like-particles \cite{2011JCAP...01..015G},
Q-balls \cite{1998PhLB..418...46K}\cite{2006hep.ex....2036B}, neutral
strangelets \cite{2006astro.ph.12740M} etc.) emitted by high-energy
sources like supernovae remnants \cite{2010APh....33..160C} or microquasars
\cite{2010MNRAS.407.2468Z}. In principle, also transient phenomena
like supernovae or binary star collapse could be interesting sources;
however non-relativistic particles will arrive scattered for thousands
or million years after their optical counterpart. Only if the flux
is sufficiently large, could one still detect alignments of slow neutral
particles with just slightly different velocities with arrivals spaced
by days or months. In this paper we do not consider transient phenomena
and time delays and assume that a sufficient and continuous flux of
particles reaches the Earth.

Two bodies could represent useful lenses: the Sun and the galactic
halo. As we will show below, the Sun in quadrature with respect to
the source will deflect SMaNPs above 0.1 degrees if they are slower
than 1500 km/sec (in the observer rest frame). Our galactic halo can
deflect particles of the same velocity or smaller up to several degrees
and its therefore the primary choice for the lens. Here ``deflection''
means in fact ``deflection with respect to the optical or relativistic
counterpart'' of the source, or ``differential deflection'' for
short. The angle of deflection $\hat{\alpha}$ itself is in fact not
observable since in general we do not know the unlensed source position.
If the galactic halo acts as a lens, and assuming radial symmetry
of the halo potential, the detection pattern would show up as straight
alignments on the sky pointing toward the galactic center, as illustrated
qualitatively in Fig. (\ref{fig:Angles-2}). Extragalactic sources
and lenses are also possible but we will not consider them in this
paper.

There are three obvious problems with this technique. The angular
resolution for non-relativistic particles is currently very poor:
directional dark matter detectors exploit the fact that the deviation
between the recoil direction of the detector's nuclei and the arrival
direction is related to the recoil energy but for non-relativistic
particles the resolution of current and planned detectors is very
low, no better than 15 degrees \cite{2010IJMPA..25....1A,2009MPLA...24.1793S}.
Cosmic ray experiments and neutrino detectors reach much higher resolutions,
down to 0.1 degrees, but only for ultra-relativistic particles. In
this paper we assume optimistically that the resolution of 1 degree
can be reached in the near future also for non-relativistic scattering.
The lower the resolution, the more difficult is to distinguish the
lensing alignment from chance alignment of background sources, halo
dark matter or other unlensed particles.

The second problem is that the lensing technique is sensitive to an
extremely narrow energy band. Typically, if the energy is larger than
the particle mass by more than $10^{-5}$ or so, the deflection by
the galactic halo reduces to the sub-degree level and cannot be resolved
by current or foreseeable particle telescopes. In fact, since they
must be non-relativistic, all detectable lensed particles of the same
species should have almost exactly the same energy, practically coinciding
with their mass. This reduces drastically the flux of non-relativistically
lensed particles but on the other hand could simplify their search,
since only equal-energy detections could have been aligned due to
particle lensing (this of course applies only if the particles transfer
all, or a fixed fraction, of their energy to the detector).

The third problem is that, if the source is moving with respect to
the observer, the non-relativistic particles that reach today the
Earth were emitted when the source was in another position. This induces
a spread of the arrival directions with respect to the relativistic
counterpart. The particles will then appear aligned along the sky
projection of the source trajectory (``proper motion vector'').
This ``proper motion alignment'' (PMA) adds to the lensing alignment
and in general produces a new alignment intermediate between the source-lens
one and the source proper motion vector. If the source proper motion
is known, the PMA can be disentangled from the lensing effect. This
paper is devoted to the calculation of the lensing alignment; we confine
in an Appendix a brief but sufficient treatment of the PMA. In general,
the proper motion does not spoil the lensing signal and in principle
can be accounted for.

In this paper we derive the geodesic equation in a general-relativistic
setting, bridging the gap between Eq. (\ref{eq:rlens}) and (\ref{eq:nrlens})
and we derive the angle of deflection as a function of velocity or
of energy per mass. Although as we already emphasized we cannot at
the moment point to any specific realistic target, the method we propose
is so straightforward to carry out on existing and future detection
maps that we feel it is an interesting addition to the astroparticle
physics toolbox.

\section{Geodesic equation}

We begin by writing down the equations of motion in the weak and slow-varying
field limit but without limitations on velocities. We assume the usual
longitudinal metric for scalar perturbations in a Friedmann-Robertson-Walker
\begin{equation}
ds^{2}=-(1+2\Psi)dt^{2}+a^{2}(1+2\Phi)dx^{i}dx_{i}
\end{equation}
 The passage to a Minkowsky metric, appropriate for Galaxy lensing,
will be carried out at the end. Defining the affine parameter $\tau$
such that $d\tau^{2}=-ds^{2}$ we can write 
\begin{equation}
d\tau=dt(1+\Psi)\Gamma
\end{equation}
 where for small $\Phi,\Psi$, 
\[
\Gamma\equiv[1-v^{2}(1+2(\Phi-\Psi))]^{1/2}
\]
 and where we define the peculiar velocity 
\begin{equation}
v^{i}\equiv a\frac{dx^{i}}{dt}
\end{equation}
 We need to derive the geodesic equations for a particle 
\begin{equation}
\frac{du^{\mu}}{d\tau}+\Gamma_{\alpha\beta}^{\mu}u^{\alpha}u^{\beta}=0
\end{equation}
 where the four velocity is 
\begin{equation}
u^{\alpha}=\frac{dx^{\alpha}}{d\tau}=\{\frac{1-\Psi}{\Gamma},\frac{v^{i}}{a\Gamma}\}
\end{equation}
 Here we assume that the gravitational potentials are small, but the
velocity can be relativistic. We assume from now on that the time
derivatives of $\Psi,\Phi$ are negligible, i.e. that the lens is
a static object. Notice that $d/dt$ is a total derivative and therefore
\begin{equation}
\frac{d(v^{i}(1-\Psi)/\Gamma)}{dt}\approx\frac{1}{\Gamma}(\dot{v}^{i}+v^{i}\frac{v\dot{v}}{\Gamma^{2}}-\frac{v^{i}v^{j}\Psi_{,j}-v^{2}v^{i}v^{j}\Phi_{,j}}{a\Gamma^{2}})
\end{equation}
 where $v=|\mathbf{v}|$ .

From the geodesic equation with $\mu=i$ we obtain to first order
in $\Phi,\Psi$:

\begin{equation}
a\dot{v_{i}}(1+\varepsilon_{1})+\gamma^{2}av_{i}v\dot{v}(1+\varepsilon_{2})=-aHv_{i}(1+\varepsilon_{1})+v^{2}\Phi_{,i}-(\gamma^{2}+1)v_{i}(v_{j}\nabla_{j}\Phi)+\gamma^{2}v_{i}(v_{j}\nabla_{j}\Psi)-(\Psi_{,i}-v^{2}\Phi_{,i})
\end{equation}
 with 
\begin{equation}
\varepsilon_{n}=2\Psi(1-2n\gamma^{2})
\end{equation}
 and where we used the usual special-relativistic factor 
\begin{equation}
\gamma^{2}=(1-v^{2})^{-1}
\end{equation}
 We put now $\Phi=-\Psi$, valid for ordinary matter. Then we obtain
\begin{equation}
\dot{v_{i}}(1+\varepsilon_{1})+\gamma^{2}v_{i}v\dot{v}(1+\varepsilon_{2})=-Hv_{i}(1+\varepsilon_{1})+(2\gamma^{2}+1)v_{i}(v_{j}\nabla_{j}\Psi)-(1+v^{2})\Psi_{,i}\label{eq:spatialnocoup}
\end{equation}
 and from now on all spatial derivatives are with respect to physical
distances $r_{i}=ax_{i}$. The $\mu=0$ equation is 
\begin{equation}
\gamma^{2}v\dot{v}(1+\varepsilon_{2})=-v^{2}H(1+\varepsilon_{1}-4\Psi)+(2\gamma^{2}-3)(v_{j}\nabla_{j}\Psi)
\end{equation}
 (this can also be obtained up to terms higher order in $\Psi$ by
multiplying Eq. (\ref{eq:spatialnocoup}) by $v^{i}$) and we can
use it to simplify Eq. (\ref{eq:spatialnocoup}) obtaining 
\begin{equation}
\gamma^{2}\dot{\mathbf{v}}(1+\varepsilon_{1})=-H\mathbf{v}(1-2\Psi)+4\gamma^{2}\mathbf{v}(\mathbf{v}\cdot\mathbf{\nabla}\Psi)-(1+2v^{2}\gamma^{2})\mathbf{\nabla}\Psi
\end{equation}

However all the equations here assume that $\gamma^{2}\Psi\ll1$,
otherwise we would need to include terms like $\gamma^{4}\Psi^{2}$
etc.. In this limit we can approximate $\hat{\varepsilon}_{1}\approx0$
so 
\begin{equation}
\gamma^{2}\dot{\mathbf{v}}=-\tilde{H}\mathbf{v}+\gamma^{2}[2\mathbf{v}\mathbf{v}\cdot\nabla(\Psi-\Phi)-v^{2}\nabla(\Psi-\Phi)]-\nabla\Psi\label{eq:master}
\end{equation}
 In Appendix A we generalize this equation to a scalar-tensor gravity.
Although the limit $v\to1$ cannot be naively taken from this equation,
it turns out that it approximates the correct result. In fact, if
a light ray passes near a spherical potential $\Phi$, in the vicinity
of the potential (where the effect is largest), the direction is orthogonal
to the gradient and therefore $\mathbf{v}\cdot\nabla(\Psi-\Phi)\to0$
and for $\gamma\to\infty$ 
\begin{equation}
\dot{\mathbf{v}}=-v^{2}\nabla(\Psi-\Phi)=-\nabla(\Psi-\Phi)\label{eq:gvar-1-2-2-2-2-1}
\end{equation}
 which is indeed the lensing equation (see e.g. \cite{2003moco.book.....D}).

\section{\label{sec:Particle-lensing}Particle lensing}

We evaluate now the lensing of relativistic and non-relativistic particles.
From here on, we restrict to a Minkowky metric, i.e. we put $H=0$
and as before we put $\Phi=-\Psi$. Eq. (\ref{eq:master}) becomes
\begin{equation}
\gamma^{2}\dot{\mathbf{v}}=-2\gamma^{2}[2\mathbf{v}\mathbf{v}\cdot\nabla\Phi-v^{2}\nabla\Phi]+\nabla\Phi\label{slowlens}
\end{equation}
 We integrate this equation along a slightly distorted path. We can
assume that the real trajectory is only weakly distorted with respect
to the imperturbed one, which is given by 
\begin{equation}
vdt=vad\eta=adr
\end{equation}
 with $v=const$. Then on the rhs of Eq. (\ref{slowlens}) we can
assume that $\mathbf{v}$ follows the imperturbed path. We consider
a point-like mass (lens) located at a distance $r_{L}$ from the observer
(see Fig. \ref{fig:Angles-1}). We have then the standard Newtonian
potential 
\begin{equation}
\Phi(r)=-G\int\frac{d^{3}r'}{|\mathbf{r}-\mathbf{r}'|}\rho(\mathbf{r}')
\end{equation}

We use a frame such that the particle is described by the coordinates
$(r\theta_{1},r\theta_{2},r)$ with small deviation angles $\theta_{1,2}$.
On this path, the term $\mathbf{v}\cdot\nabla\Phi$ changes sign when
the particle overcomes the lens and therefore its total effect for
a spherically symmetric lens is zero. We are then left with ($i=1,2$)
\begin{equation}
\dot{v}_{i}=v^{2}\frac{d(r\theta_{i})}{adr^{2}}=2v^{2}\nabla_{i}\Phi+\gamma^{-2}\nabla_{i}\Phi\label{slowlens-1}
\end{equation}
 or 
\begin{equation}
\frac{d(r\theta_{i})}{dr^{2}}=2\nabla_{i}\Phi+\frac{1}{v^{2}\gamma^{2}}\nabla_{i}\Phi\label{slowlens-1-1}
\end{equation}
 where now spatial derivatives are with respect to comoving coordinates.
The total lensing can be obtained by summing the two terms on the
rhs. Let us analyse them separately.

The first gives the standard weak-lensing equation. Its solution is
(see eg \cite{2003moco.book.....D}) 
\begin{equation}
\alpha^{i}=\theta^{i}-\theta_{0}^{i}=2\int_{0}^{r_{s}}dr\frac{r_{s}-r}{r_{s}r}\frac{d\Phi}{d\theta^{i}}\equiv\nabla_{\perp i}\phi_{1}
\end{equation}
 where $\nabla_{\perp}$ is the gradient along the angular directions
$\theta_{1,2}$ , $r_{s}$ is the source distance and $\phi_{1}$
is defined as 
\begin{eqnarray}
\phi_{1} & = & 2\int_{0}^{r_{s}}dr\frac{r_{s}-r}{r_{s}r}\Phi=-2G\int_{0}^{r_{s}}dr\frac{r_{s}-r}{r_{s}r}\int\frac{d^{3}r'}{|\mathbf{r}-\mathbf{r}'|}\rho(\mathbf{r}')\label{eq:phi1}\\
 & = & -2G\frac{r_{s}-r_{L}}{r_{s}r_{L}}\int\frac{d^{2}R'dr'dr}{((\mathbf{R'}-\mathbf{R})^{2}+(r-r')^{2})^{1/2}}\rho(\mathbf{R'},r')
\end{eqnarray}
 where $\mathbf{R}'=(r'\theta'_{1},r'\theta'_{2})$ and $\mathbf{R}=(r_{L}\theta{}_{1},r_{L}\theta{}_{2})$
and where if the lens potential is sharply peaked at the lens position
we can assume $r=r_{L}$ in the slowly varying factors inside the
integral (in other words we assume here that the lens is a point mass
separated by the source by a small angle, i.e. the typical strong-lensing
astronomical configuration; later on we generalize to a distributed
mass). Finally we have (up to a constant independent of $\theta$
and therefore irrelevant when the gradient $\nabla_{\perp}$ is applied)
\begin{equation}
\phi_{1}(\mathbf{\theta})=4G\frac{r_{s}-r_{L}}{r_{s}r_{L}}\int d^{2}R'\Sigma(\mathbf{R'})\ln|\mathbf{R'}-\mathbf{R}|
\end{equation}
 where the surface mass density is 
\begin{equation}
\Sigma=\int dr\rho(\mathbf{R},r)
\end{equation}
 The second term in (\ref{slowlens-1-1}) can be treated similarly
and gives 
\begin{equation}
\phi_{2}(\mathbf{\theta})=2G\frac{r_{s}-r_{L}}{v^{2}\gamma^{2}r_{s}r_{L}}\int d^{2}R'\Sigma(\mathbf{R'})\ln|\mathbf{R'}-\mathbf{R}|
\end{equation}
 The second term is therefore $(2v^{2}\gamma^{2})^{-1}$ times the
first one, and dominates for $v\ll1$. The final result is 
\begin{equation}
\mathbf{\alpha}=\mathbf{\nabla}_{\perp}\phi
\end{equation}
 where 
\begin{equation}
\phi=4G(1+\frac{1}{2v^{2}\gamma^{2}})\frac{r_{s}-r_{L}}{r_{s}r_{L}}\int d^{2}R'\Sigma(\mathbf{R'})\ln|\mathbf{R'}-\mathbf{R}|
\end{equation}
 So finally we have 
\begin{equation}
\mathbf{\alpha}(\boldsymbol{\mathbf{\mathbf{\xi}}})=4G(1+\frac{1}{2v^{2}\gamma^{2}})\frac{r_{s}-r_{L}}{r_{s}}\int d^{2}\xi'\frac{\Sigma(\mathbf{\boldsymbol{\mathbf{\mathbf{\xi}}}'})}{|\boldsymbol{\mathbf{\mathbf{\xi}}}-\boldsymbol{\mathbf{\mathbf{\xi}}}'|^{2}}(\boldsymbol{\mathbf{\mathbf{\xi}}}-\boldsymbol{\mathbf{\mathbf{\xi}}}')
\end{equation}
 where $\boldsymbol{\mathbf{\mathbf{\mathbf{\xi}}}}$ is a vector
on the lens plane (impact vector). For a point lens of mass $M$,
this gives 
\begin{equation}
\mathbf{\alpha}=4GM(1+\frac{1}{2v^{2}\gamma^{2}})\frac{r_{s}-r_{L}}{r_{s}\xi}
\end{equation}
 where $\xi=|\boldsymbol{\mathbf{\mathbf{\xi}}}|=r_{L}\sin\delta$
is the impact radius. The angle $\alpha$ is related to the angle
$\hat{\alpha}$ (see Fig. \ref{fig:Angles-1}) between initial and
final propagation (more often employed by astronomers) by the relation
\begin{equation}
\hat{\alpha}=\alpha\frac{r_{s}}{r_{s}-r_{L}}=4GM(1+\frac{1}{2v^{2}\gamma^{2}})\frac{1}{\xi}=\frac{2\Pi R_{S}}{\xi}\label{eq:genalpha}
\end{equation}
 where $R_{S}=2GM/c^{2}$ is the Schwarzschild radius and 
\begin{equation}
\Pi=1+\frac{1}{2v^{2}\gamma^{2}}
\end{equation}
 Eq. (\ref{eq:genalpha}) generalizes Eqs. (\ref{eq:rlens},\ref{eq:nrlens}).

Using these formulae, we can employ the standard lensing equations,
simply replacing $\Pi R_{S}$ for $R_{S}$. Making reference to Fig.
(\ref{fig:Angles-1}), for small $\delta$, the imperturbed angle
between source and lens, the angular separation between the direction
of the lens and the direction of particle arrival is (see eg. \cite{1992grle.book.....S})
\begin{equation}
\theta_{1,2}=\frac{1}{2}(\delta\pm\sqrt{4\tau^{2}+\delta^{2}})
\end{equation}
 where 
\begin{equation}
\tau=\Pi^{1/2}\sqrt{2R_{S}\frac{r_{s}-r_{L}}{r_{L}r_{s}}}\equiv\Pi^{1/2}\alpha_{0}
\end{equation}
 and where $\alpha_{0}$ is the standard characteristic angle. Finally,
the differential deflection, i.e. the angle between the different
arrival directions of particles of velocity $v$ with respect to the
optical/relativistic counterpart will be 
\begin{equation}
\Delta\theta=\frac{1}{2}(\sqrt{4\Pi\alpha_{0}^{2}+\delta^{2}}-\sqrt{4\alpha_{0}^{2}+\delta^{2}})\label{eq:oldlenssol}
\end{equation}
 where we only considered the solution with $\theta>\delta$ . From
this equation we can identify three regimes of velocities. For very
small $v$, the dominating term is $4\Pi\alpha_{0}^{2}$ and we expect
a $1/v$ behavior: 
\begin{equation}
\Delta\theta\approx\Pi^{1/2}\alpha_{0}\approx\frac{1}{v}\sqrt{R_{S}\frac{r_{s}-r_{L}}{r_{L}r_{S}}}\to\frac{1}{v}\sqrt{\frac{R_{S}}{r_{L}}}\label{eq:strong}
\end{equation}
 (where in the last expression, here and in the next equation, we
assume $r_{s}\gg r_{L}$). For larger $v$, but still $v\ll1$, we
expect a $1/v^{2}$ trend since 
\begin{equation}
\Delta\theta\approx\frac{\alpha_{0}^{2}}{\delta}(\Pi-1)=\frac{\alpha_{0}^{2}}{\delta}\frac{1}{2v^{2}\gamma^{2}}\to\frac{R_{S}}{r_{L}\delta v^{2}}\label{eq:strong2}
\end{equation}
 Finally, for $v\to1$ , $\Delta\theta$ quickly vanishes.

\section{\label{sec:Lensing-in-a}Lensing in a halo}

In the following we derive the lensing in our own galaxy halo, where
the approximation of point mass and of small $\delta$ is no longer
acceptable. We assume a general spherical mass distribution $M(r)$
centered on $\mathbf{r}_{L}=(r_{L}\cos\theta,r_{L}\sin\theta)$ on
the propagation plane. In Fig. (\ref{fig:Angles-1}) we display the
geometric configuration of the problem. In this case the lensing equation
amounts to the statement that $S'S+SC=S'C$ , from which, assuming
that OA$\approx$$r_{L}\cos\delta$ and OS'$\approx$OS, one finds
to first order in $\hat{\alpha}$ 
\begin{equation}
\tan\theta=\tan\delta+\hat{\alpha}\frac{r_{s}-r_{L}\cos\delta}{r_{s}}
\end{equation}
 where if $\Phi$ is the gravitational potential of the halo (see
e.g. Eq. (\ref{eq:phi1})) 
\begin{eqnarray}
\hat{\mathbf{\alpha}}(\delta,v) & = & 2\Pi\frac{r_{s}}{r_{s}-r_{L}\cos\delta}\int_{0}^{r_{s}}dr\frac{r_{s}-r}{r_{s}r}\frac{d}{d\eta}\Phi(|\mathbf{r}-\mathbf{r}_{L}|)|_{\eta=0}\\
 & = & 2\Pi\frac{\hat{R_{s}}}{r_{L}\sin\theta}\left[\frac{1}{2}\int_{0}^{r_{s}}dr\frac{r_{s}-r}{r_{s}-r_{L}\cos\delta}\frac{r_{L}\sin\delta}{r\hat{M}}\frac{d}{d\eta}\Phi(|\mathbf{r}-\mathbf{r}_{L}|)|_{\eta=0}\right]\label{eq:haloa}
\end{eqnarray}
 where $\hat{R_{s}}=2G\hat{M}$ and $\hat{M}=M(r_{L}\sin\delta)$
and the distance between the lens center and a generic point $\mathbf{r}=(r\cos\eta,r\sin\eta)$
is 
\begin{equation}
|\mathbf{r}-\mathbf{r}_{L}|=[r^{2}+r_{L}^{2}-2rr_{L}\cos(\eta-\delta)]^{1/2}
\end{equation}
 Notice that $\hat{R}_{S}$ is the Schwarzschild radius of the mass
contained inside the impact radius. Notice also that here we put the
coordinates of the center as $\mathbf{r}_{L}=(r_{L}\sin\delta,r_{L}\cos\delta)$,
since at this order $\theta$ can be replaced by $\delta$. To the
same order, we can approximate the particle trajectory with the horizontal
straight line $\eta=0$ (i.e. $y=0$ ). The dimensionless term inside
the square brackets (let us call it the halo integral) 
\begin{equation}
H\equiv\frac{1}{2}\int_{0}^{r_{s}}dr\frac{r_{s}-r}{r_{s}-r_{L}\cos\delta}\frac{r_{L}\sin\delta}{r\hat{M}}\frac{d}{d\eta}\Phi(|\mathbf{r}-\mathbf{r}_{L}|)|_{\eta=0}\label{eq:halo}
\end{equation}
 reduces to unity for $M=const$ and for small $\delta$. We have
infact in this case $\Phi(x)=\hat{M}/x$ and 
\begin{eqnarray}
H & \equiv & \frac{1}{2}\int_{0}^{r_{s}}dr\frac{r_{s}-r}{r_{s}-r_{L}\cos\delta}\frac{r_{L}\sin\delta}{r}\frac{d}{d\eta}\left(\frac{1}{|\mathbf{r}-\mathbf{r}_{L}|}\right)|_{\eta=0}\\
 & = & \frac{1}{2}\int_{0}^{r_{s}}dr\left(\frac{r_{s}-r}{r_{s}-r_{L}\cos\delta}\right)\frac{(r_{L}\sin\delta)^{2}}{(r^{2}+r_{L}^{2}-2rr_{L}\cos\delta)^{3/2}}\label{eq:h3}
\end{eqnarray}
 The integrand is sharply peaked at $r\approx r_{L}\cos\delta$ and
therefore we can replace $r$ with $r_{L}\cos\delta$ in the first
parentheses of Eq. (\ref{eq:h3}) and find, in the limit of $r_{s}\gg r_{L}$,
\begin{equation}
\frac{1}{2}\int_{0}^{r_{s}}dr\frac{(r_{L}\sin\delta)^{2}}{(r^{2}+r_{L}^{2}-2rr_{L}\cos\delta)^{3/2}}=\frac{1+\cos\delta}{2}\approx1
\end{equation}
 (the condition $r_{s}\gg r_{L}$ can in fact be relaxed; for instance,
$H$ approximates 1 even in the isosceles configuration $r_{s}=2r_{L}\cos\delta$).

Since we have now $\hat{\alpha}=2\Pi H\hat{R}_{S}/(r_{L}\sin\theta)$
we can solve the lens equation 
\begin{equation}
\tan\theta=\tan\delta+\frac{\Pi H}{\sin\theta}\label{eq:newlenseq}
\end{equation}
 where 
\begin{equation}
\alpha_{0}^{2}=\frac{2\hat{R}_{S}(r_{s}-r_{L}\cos\delta)}{r_{s}r_{L}}
\end{equation}
 is a minor generalization of the standard characteristic angle. In
Eq. (\ref{eq:newlenseq}) the factor $\Pi$ takes into account the
velocity of the particles, the factor $H$ takes into account the
deviation from the point mass configuration. Both reduce to unity
for the standard point-mass relativistic lensing. Although Eq. (\ref{eq:newlenseq})
could be easily solved numerically, here we assume simply that $\cos\theta\approx1$
even when $x$ is not very small, and obtain the solutions 
\begin{equation}
\theta_{1,2}=\arcsin\left[\frac{1}{2}(\tan\delta\pm\sqrt{4\tau^{2}+\tan^{2}\delta})\right]
\end{equation}
 where now 
\begin{equation}
\tau^{2}=\Pi H\alpha_{0}^{2}
\end{equation}
 Finally, the halo differential deflection is 
\begin{equation}
\Delta\theta=\arcsin\left[\frac{1}{2}(\tan\delta+\sqrt{4\Pi H\alpha_{0}^{2}+\tan^{2}\delta})\right]-\arcsin\left[\frac{1}{2}(\tan\delta+\sqrt{4H\alpha_{0}^{2}+\tan^{2}\delta})\right]\label{eq:deltathetafin}
\end{equation}
 which generalizes Eq. (\ref{eq:oldlenssol}). Notice that for $\tan\delta\ge1$
the argument of the $\arcsin$ exceeds unity; in this case one should
solve the full Eq. (\ref{eq:newlenseq}). If $H=1$ and moreover $\delta\ll1$
(along with $\alpha_{0}\ll1$, an approximation needed in both cases),
we recover Eq. (\ref{eq:oldlenssol}). The same three regimes of velocities
discussed before can be found here, although with more complicated
behaviors.

\section{Sun and star lensing}

For a star of 1 solar mass with $r_{s}\gg r_{L}$ one has 
\begin{equation}
\alpha_{0}^{2}=\frac{2R_{S}}{r_{L}}\approx10^{-5}\left(\frac{1kpc}{r_{L}}\right)\mathrm{arcsec}^{2}
\end{equation}
 For the Sun at $r_{L}=1$AU we have then, using Eq. (\ref{eq:strong2})
\begin{equation}
\Delta\theta_{sun}=\frac{10^{3}}{\delta_{sun}}\frac{1}{v^{2}}\mathrm{arcsec}
\end{equation}
 so that for a source at $\delta=1$ rad the deflection is 
\begin{equation}
\Delta\theta_{sun}\approx\frac{10^{-2}}{4v^{2}}\mathrm{arcsec}\label{eq:defsun}
\end{equation}
 We will see in the next section that the solar deflection, although
non negligible, is quite smaller than the halo deflection. Moreover,
it can be evaluated and subtracted from the total deflection since
we know position and mass of the Sun with great accuracy.

Assuming a density of $10^{-1}$ stars per parsec$^{3}$, in a cone
of 1 arcsec aperture and 10 kpc depth, one expect roughly one star.
The deflection for particles passing 1 arcsec away from a 1 solar
mass star lens at 1 kpc is 
\begin{equation}
\Delta\theta_{star}=\frac{10^{-5}}{\delta_{star}}\frac{1}{2v^{2}\gamma^{2}}\mathrm{arcsec}\approx\frac{10^{-5}}{2v^{2}}\mathrm{arcsec}
\end{equation}
 again using using Eq. (\ref{eq:strong2}). The deflection from single
stars is therefore negligible with respect to the solar one. To estimate
the resulting effect of many stars we can integrate the random deflection
variance $(\Delta\theta_{star})^{2}$ over a cone of aperture $\theta_{max}$
(expressed in arcseconds) along whose axis the particle travels from
the source to us. Then we have 
\begin{equation}
(\Delta\theta_{tot})^{2}=\frac{2\pi}{C^{2}}\int_{0}^{r_{s}}dr\int_{\theta_{min}}^{\theta_{max}}d\theta\left(\frac{R_{s}^{2}C^{2}}{v^{2}\theta}\right)^{2}\left(\frac{r_{s}-r}{r_{s}r}\right)^{2}n_{s}r^{2}\theta
\end{equation}
 where $C\approx2\cdot10^{5}$ converts from radians to arcsec and
$n_{s}=0.1$pc$^{-3}$ is the star density. Then we have 
\begin{equation}
(\Delta\theta_{tot})^{2}=\frac{2\pi}{C^{2}}n_{s}\frac{r_{s}}{3}\left(\frac{R_{s}^{2}C^{2}}{v^{2}\theta}\right)^{2}\log\frac{\theta_{max}}{\theta_{min}}\approx\left(\frac{3\cdot10^{-6}}{v^{2}}\right)^{2}\mathrm{arcsec}^{2}
\end{equation}
 where in the last step we assume, for instance, $\theta_{max}=10$
deg and $\theta_{min}=1$ arcsec. The total random deflection induced
by the stars along the line-of-sight is therefore negligible as well.

\section{Galaxy halo deflection}

Let us consider now the amount of deflection induced by the Milky
Way halo potential. We assume that a source lies at distance $r_{s}$
from us in a direction that makes an angle $\delta$ with respect
to the Galactic center (GC). The GC is then at coordinates $(R_{c}\cos\delta,R_{c}\sin\delta)$
on the plane observer-GC-source if $R_{c}$ is the GC distance from
us.

As a first approximation we can simply take Eq. (\ref{eq:strong2})
and put $M=10^{10}M_{\odot}$. In this case the characteristic angle
is of the order of 
\begin{equation}
\alpha_{0}^{2}\approx10^{5}\left(\frac{1kpc}{r_{L}}\right)\left(\frac{M}{10^{10}M_{\odot}}\right)arcsec^{2}
\end{equation}
 For $v\ll1$ we have from Eq. (\ref{eq:strong2}) 
\begin{equation}
\Delta\theta_{gal}=\frac{\alpha_{0}^{2}}{\delta}\frac{1}{2v^{2}\gamma^{2}}\approx\frac{10^{5}}{2\delta v^{2}}(\frac{1kpc}{r_{L}})arcsec
\end{equation}
 Assuming $\delta=0.25$ rad and $r_{L}=10$ kpc this is $\approx0.1/v^{2}$
arcsec. As anticipated, this is one order of magnitude larger than
Sun's deflection in Eq. (\ref{eq:defsun}). This is larger than 1
deg for $v<0.005$. For $v\ll0.001$ the deflection is of the order
of tens of degrees and the small-deflection approximation adopted
throughout would fail. 

For a more precise calculation, we need to integrate Eq. (\ref{eq:haloa})
with $\mathbf{r}_{L}=(R_{c}\cos\delta,R_{c}\sin\delta)$. The mass
$M(x)$ inside a distance $x$ from the center can be taken by integrating
a model for the Galaxy. Here, just to produce a rough estimate, we
employ a Navarro-Frenk-White \cite{1996ApJ...462..563N} profile for
the galatic halo 
\begin{equation}
\rho_{NFW}(r)=\frac{x_{s}^{3}\rho_{0}}{r(x_{s}+r)^{2}}
\end{equation}
 with parameters $x_{s}\equiv r_{v}/c$ where $r_{v}$ is the virial
radius and $c$ is the concentration. For our source we choose $c=12.5$,
$r_{v}=200$ kpc corresponding to $x_{s}=16$ kpc. The density parameter
is $\rho_{0}=0.014M_{\odot}/pc^{3}$ is chosen such that the mass
within the virial radius is $M=10^{12}M_{\odot}$ \cite{2008PhLB..660...81A,2008MNRAS.389.1911S}.

\begin{table}
\begin{tabular}{|c|c|c|c|}
\hline 
Parameter  & Halo  & Sun  & Star\tabularnewline
\hline 
\hline 
$M$  & $M_{NFW}$  & $M_{\odot}$  & $M_{\odot}$\tabularnewline
\hline 
$r_{L}$  & $R_{c}$  & 1 A.U.  & $1$ kpc\tabularnewline
\hline 
$r_{s}$  & $2r_{L}$cos $\delta$  & 10 kpc  & $\gg r_{L}$ \tabularnewline
\hline 
\end{tabular}

\caption{Parameters used for the Milky Way NFW halo, sun and a typical Milky
Way star. The distance of the Galactic Center from Earth is $R_{c}=8.5$
kpc. $M{}_{NFW}(x)$ is calculated from the integral (\ref{eq:halo})
as a function of the distance $r$ from the center of the lens. Alternatively,
for the Galaxy Halo, we also compare with a constant mass lens of
mass $M=M(r_{L}sin\delta)$.}

\label{cosmological_parameters} 
\end{table}

\begin{figure}
\includegraphics[scale=0.4]{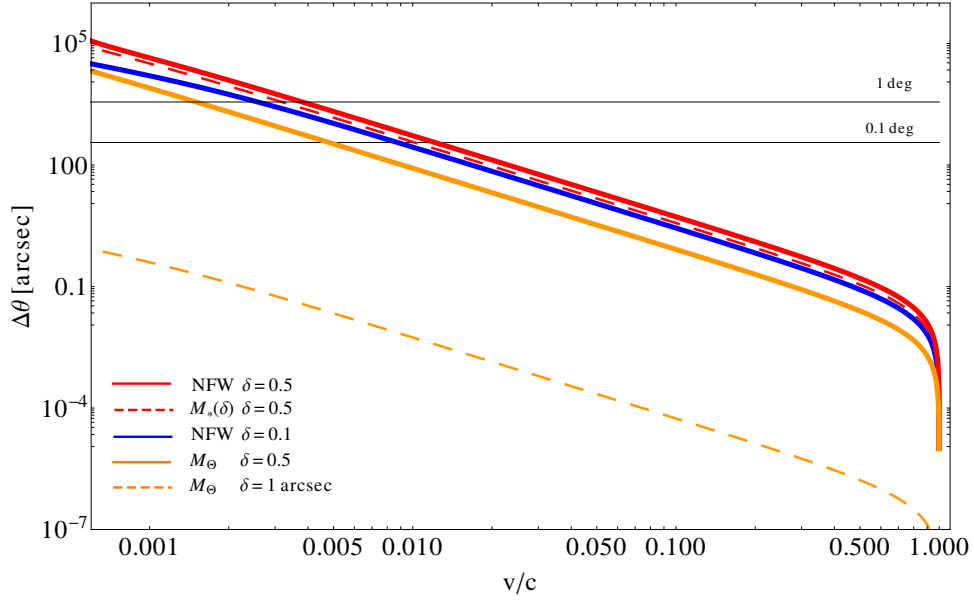}

\caption{\label{fig:Differential-deflection-versus}Differential deflection
$\Delta\theta$ in units of arcsec as a function of velocity, in units
of the speed of light, for Milky Way halo, Sun and star. We show the
$\Delta\theta$ corresponding to $\delta=0.5$ rad for a NFW potential
$\phi(x)$ (solid red) and for $\delta=0.1$ (solid blue). For $\delta=0.5$
we plot also the curve assuming a constant mass $M_{*}=M(r_{L}sin\delta)$
(dashed red). The curve for $\delta=0.1$ shows clearly the three
regimes $v^{-1}$,$v^{-2}$ and $\gamma^{-1}$, from small to large
velocities. The Sun (solid orange) is shown for $\delta=0.5$ rad.
A typical star in the Milky Way of mass $M=M_{\odot}$, located at
$r_{L}=1$kpc and $r_{s}\gg r_{L}$ with $\delta=1$ arcsec is also
shown (dashed orange). Resolution thresholds of $1$ deg and $0.1$
deg are overplotted for reference (solid black).}
\end{figure}

\begin{figure}
\includegraphics[scale=0.6]{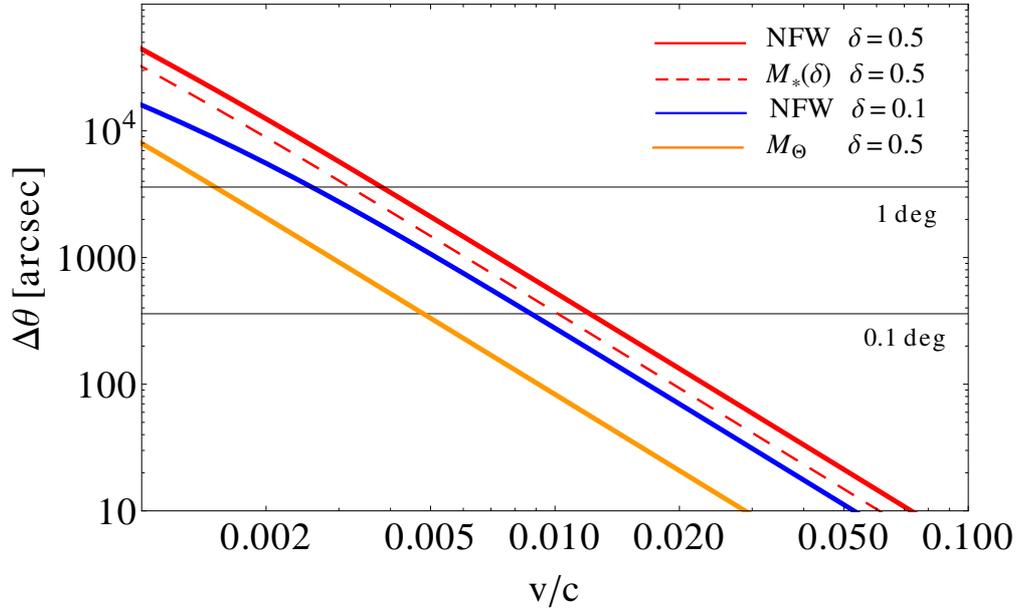}

\caption{\label{fig:Zoom-of-Fig.}Zoom of Fig. (\ref{fig:Differential-deflection-versus}).}
\end{figure}

\begin{figure}
\includegraphics[scale=0.45]{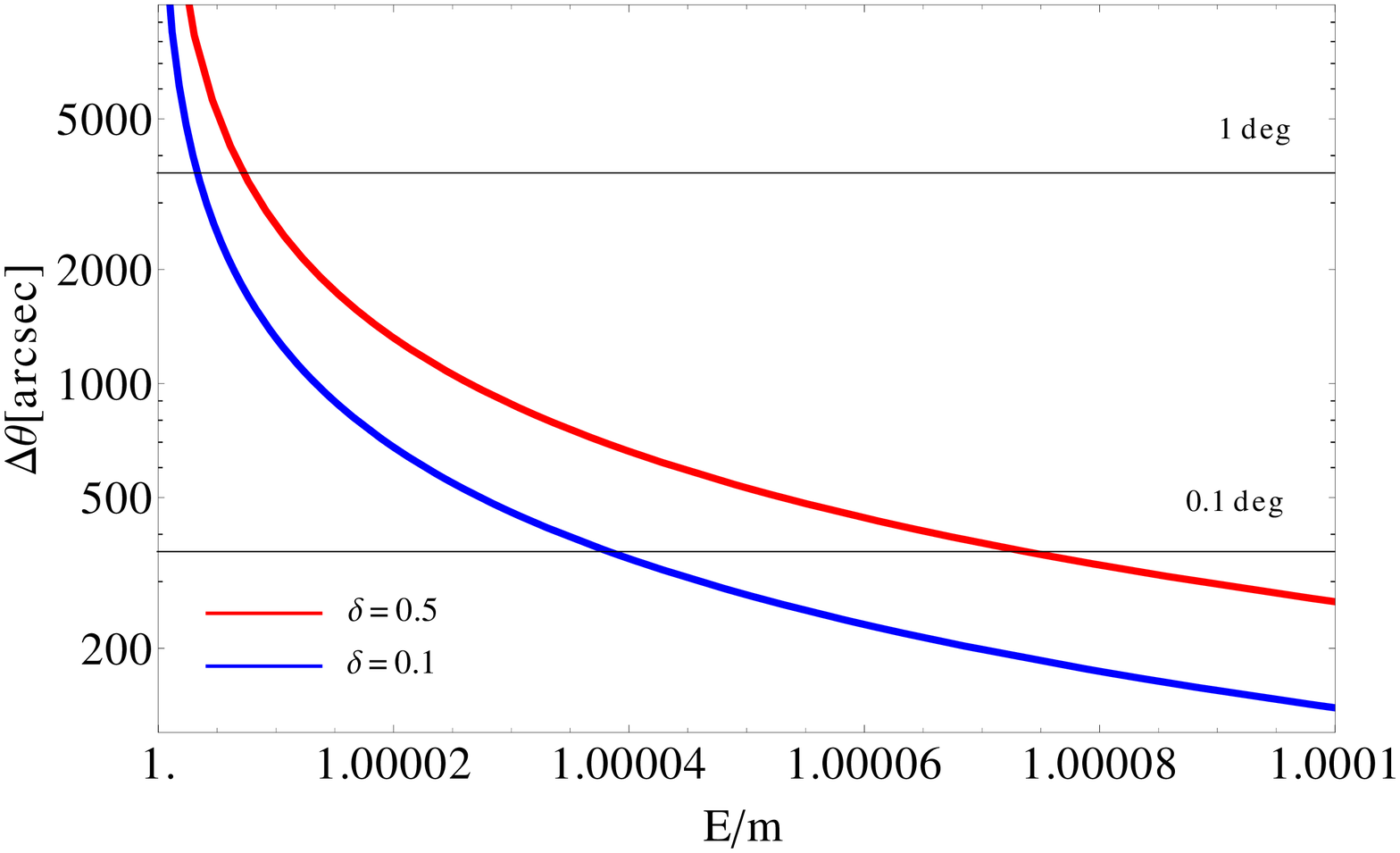}

\caption{\label{fig:Differential-deflection-versus-energy}Differential deflection
versus energy/mass for the NFW halo with $\delta=0.1$ and $0.5$,
same parameters as in Fig. (\ref{fig:Differential-deflection-versus-energy}).}
\end{figure}

Then we obtain 
\begin{equation}
M_{NFW}(x)=4\pi\int_{0}^{x}y^{2}\rho_{NFW}(y)dy
\end{equation}
 and the potential 
\begin{equation}
\Phi_{NFW}(x)=\frac{4\pi\rho_{0}x_{s}^{3}\log(1+x/x_{s})}{x}
\end{equation}
 Then we can evaluate the halo integral $H$ and use Eq. (\ref{eq:deltathetafin})
to calculate the differential deflection $\Delta\theta$. The results
are plotted in Figs. (\ref{fig:Differential-deflection-versus}) and
(\ref{fig:Zoom-of-Fig.}). As can be seen, the halo lensing can be
larger than 1 degree for $v<0.004$. The dependence on $\delta$ in
the range 0.1-0.5 rad is quite modest, since for small $\delta$ both
the impact radius and the halo mass within the particle trajectory
decrease. In Fig. (\ref{fig:Differential-deflection-versus-energy})
we plot the deflection versus $E/m\approx1+v^{2}/2$. This shows that
as soon as the particle energy is $10^{-5}$ larger than the particle
mass, the deflection decreases to less than 1 deg, making it impossible
to distinguish from the relativistic counterpart. The particle lensing
method is therefore sensitive only to a very narrow energy band, $\Delta E/E\approx10^{-5}$.

The assumption of a smooth NFW profile is of course very rough. A
realistic galaxy model should include at a minimum the disk and the
bulge. As long as they are both radially symmetric, the particles
will still be aligned along the GC direction (for the disk cilindrical
symmetry is sufficient, provided also the lens belongs to the disk).
Deviation from symmetry will induce deviations from the straight alignment.

\section{Conclusions}

In the last few years we have seen the beginning of particle astronomy,
i.e. the detection of particles of astrophysical origin with detectors
that are able to reconstruct the arrival direction. Present neutrino
and cosmic ray detectors already reach an angular resolution for high-energy
events of 1 degree and better \cite{2011arXiv1107.4809T,2011NIMPA.630..125P};
although the directional dark matter detectors that are currently
investigated only reach down to 15 degrees or so \cite{2010IJMPA..25....1A},
one can reasonably expect that future technology will sensibly improve
this, especially if additional science cases can be found in support
for it.

In this paper we explore the possibility of detecting stable, massive,
neutral particles (SMaNPs) by using particle telescopes. We point
out that if these particles are non-relativistic they will appear
to an observer aligned in the sky due to the combined effect of non-relativistic
lensing and of the source proper motion. This alignment is a very
distinctive signature that seems difficult to be mimicked by background
noise. The method can be applied to particles of any mass, provided
of course their kinetic energy is above the detector's threshold.
Searching for alignments in detection maps could then be a simple
and effective way to search for new physics and new astrophysics.

We derived the deflection-velocity relation due to halo lensing and
to proper motion and found that particles with velocity around 1000
km/sec are the best target for detection. We also pointed out however
that the energy bandwidth is extremely small, roughly $\Delta E/E\approx10^{-5}$
and therefore the useful flux from any given source is very much reduced.
There are several extensions of the standard model that predict the
existence of SMaNPs \cite{2009PhLB..670..281F,2011JCAP...01..015G,1998PhLB..418...46K,2006hep.ex....2036B,2006astro.ph.12740M};
how and where these particles could be emitted by astrophysical bodies
is of course a difficult problem that we leave to future work. 
\begin{acknowledgments}
We thank G. Bertone, M. Cirelli, N. Fornengo, M. Quartin, T. Schwetz-Mangold,
C. Wetterich for useful discussions. L.A. is supported by the Deutsche
Forschungsgemeinschaft through the programme TRR33 ``The Dark Universe''.
V.P. is supported by the Marie Curie IEF. This work was also supported by the young scientist SISSA grant. 
\end{acknowledgments}
\bibliographystyle{amsplain} \bibliographystyle{amsplain}
\bibliography{particlelensing}

\section*{Appendix A. Geodesics in a scalar-tensor gravity}

For generality, we can include the effect of a fifth-force mediated
by a scalar field, as in a modified gravity scenario (see e.g. \cite{AT}),
by conformally transforming the geodesic equation. If we put 
\begin{equation}
\hat{g}_{\mu\nu}=e^{2\beta\phi}g_{\mu\nu}
\end{equation}
 we obtain 
\begin{equation}
\Gamma_{\nu\lambda}^{\mu}=\hat{\Gamma}_{\nu\lambda}^{\mu}-\beta(\phi_{,\lambda}\delta_{\nu}^{\mu}+\phi_{,\nu}\delta_{\lambda}^{\mu}-\phi^{,\mu}\hat{g}_{\nu\lambda})
\end{equation}
 and 
\begin{equation}
u^{\mu}=e^{\beta\phi}\hat{u}^{\mu}
\end{equation}
 Then, by assuming also that $\varphi\equiv\delta\phi$ is taken at
first order and $\dot{\varphi}$ is negligible we have

\begin{equation}
\dot{\mathbf{v}}(1+\varepsilon_{1})+\gamma^{2}\mathbf{v}v\dot{v}(1+\varepsilon_{2})=-\tilde{H}\mathbf{v}(1+\varepsilon_{1})-\nabla(\Psi-\beta\varphi)-v^{2}\nabla(\Psi+\beta\varphi)+\mathbf{v}(\mathbf{v}\cdot\mathbf{\nabla}(\Psi+\beta\varphi))+2\gamma^{2}\mathbf{v}(\mathbf{v}\cdot\nabla\Psi)\label{eq:gvar-1-2}
\end{equation}
 where $\tilde{H}=H(1-\beta\frac{\dot{\phi}}{H})$ . This reduces
to the standard case 
\begin{equation}
\dot{\mathbf{v}}=-\tilde{H}\mathbf{v}-\nabla(\Psi-\beta\varphi)\label{stand-1}
\end{equation}
 for small velocities. The $\mu=0$ equation is 
\begin{equation}
\gamma^{2}v\dot{v}(1+\varepsilon_{2})=-\tilde{H}v^{2}(1+\varepsilon_{1}-4\Psi)+(\mathbf{v}\cdot\mathbf{\nabla}((2\gamma^{2}-3)\Psi+\beta\varphi))\label{eq:gvar-1-1-1}
\end{equation}
 which again can be used to simplify Eq. (\ref{eq:gvar-1-2}):

\begin{equation}
\gamma^{2}\dot{\mathbf{v}}(1+\varepsilon_{1})=-\tilde{H}\mathbf{v}(1-2\Psi)+4\gamma^{2}\mathbf{v}(\mathbf{v}\cdot\mathbf{\nabla}\Psi)-2\gamma^{2}v^{2}\nabla\Psi-\nabla(\Psi-\beta\varphi)\label{eq:gvar-1-2-2-1}
\end{equation}
 If we keep the two potentials distinct, we obtain 
\begin{equation}
\gamma^{2}\dot{\mathbf{v}}(1+\hat{\varepsilon}_{1})=-\tilde{H}\mathbf{v}(1-2\Psi)+\gamma^{2}[2\mathbf{v}\mathbf{v}\cdot\nabla(\Psi-\Phi)-v^{2}\nabla(\Psi-\Phi)]-\nabla(\Psi-\beta\varphi)\label{slowlens0-1}
\end{equation}
 where 
\begin{equation}
\hat{\varepsilon}_{1}=2\Phi(\gamma^{2}-1)-2\Psi\gamma^{2}
\end{equation}
 It is interesting to note that the motion depends on $\Psi-\beta\varphi$
alone when slow, but on $\psi\equiv\Phi-\Psi$ (and not on $\varphi$,
due to the conformal invariance of electromagnetism) when ultrarelativistic.

\section*{Appendix B. Flux from supernovae remnants}

We can estimate the possible order-of-magnitude flux of SMaNPs from
a typical supernovae remnant (SNR) in the following way. The $\gamma$-ray
energy flux from typical SNRs, e.g. the sources RX J0852.0-4622 and
RX J1713.7-3946, as observed by the H.E.S.S. Cerenkov detector \cite{2011A&A...531C...1A,2008ICRC....2..667L}
in an energy band $\Delta E$ is 
\begin{equation}
N_{\Delta E}=\frac{dN}{dE}\Delta E
\end{equation}
 where 
\begin{equation}
\frac{dN}{dE}\approx10^{-11}cm^{-2}s^{-1}TeV^{-1}\left(\frac{E}{1TeV}\right)^{-\Gamma}
\end{equation}
 is the differential flux in the range $100$GeV $-10$TeV and $\Gamma\approx2$.
Both sources are estimated to lie at a distance 1 kpc, with a large
uncertainty. The flux from similar SNRs at distance $D$ could be
taken therefore to be 
\begin{equation}
\frac{dN}{dE}\approx10^{-11}cm^{-2}s^{-1}TeV^{-1}\left(\frac{E}{1TeV}\right)^{-\Gamma}\left(\frac{D}{1kpc}\right)^{-2}
\end{equation}
 Let us now suppose that the flux of SMaNPs is a fraction $\epsilon$
of the flux in $\gamma$-rays. In an energy band $\Delta E/E=10^{-4}$
(which as we have seen is the typical bandwidth for particle lensing)
we have a flux 
\begin{equation}
N_{\Delta E}\approx10^{-4}\epsilon m^{-2}yr^{-1}\left(\frac{E}{1TeV}\right)^{-\Gamma+1}\left(\frac{D}{1kpc}\right)^{-2}
\end{equation}
 Then a flux larger than 1 particle in $T$ years can be achieved
if 
\begin{equation}
\left(\frac{m}{1TeV}\right)^{-\Gamma+1}\left(\frac{D}{1kpc}\right)^{-2}\epsilon AT>10^{4}
\end{equation}
 where $A$ is the detector area in m$^{2}$ . A detector of effective
area $10^{3}$m$^{2}$ (compare with 400m$^{2}$ for H.E.S.S.) will
therefore receive more than one particle/year if $m\approx0.1$TeV
and $D\approx1$kpc and $\epsilon\approx1$.

Obviously, both $\varepsilon$ and the fraction of particles effectively
detected can be much less than unity, especially if the particles
are weakly interacting; the quick estimate of this Appendix is meant
only to give an idea of the orders of magnitudes.

\section*{Appendix C. Source's proper motion}

If the source has a non-negligible proper motion with respect to us
then the non-relativistic particles that we observe today were emitted
from a position that depends on their velocity. The particles will
have therefore a different arrival direction with respect to the relativistic
counterpart. This will spread the particles on the plane that contains
the source velocity vector and the observer, inducing a ``proper
motion alignment''.

\begin{figure}
\includegraphics[bb=0bp 200bp 595bp 842bp,clip,scale=0.4]{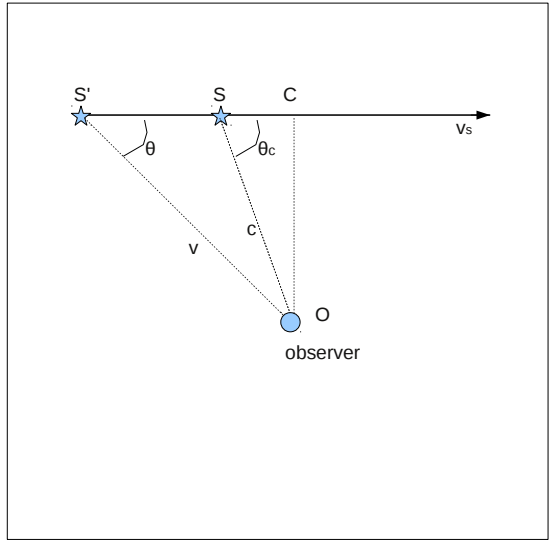}

\caption{\label{fig:Proper-motion-geometry}Proper motion geometry. The source
moves with velocity $v_{s}$ from S' to S. When the source is at S'
it emits a particle with velocity $v$; when is in S it emits light
or a ultra-relativistic particle with velocity $c$. The distance
OC is the impact radius $\xi_{PM}$. The time it takes for the particle
to travel from S' to the observer must equal the times it takes for
the source to move from S' to S and for the relativistic signal to
propagate from S to the observer.}
\end{figure}

Let us now denote with $\theta_{c}$ (see Fig. \ref{fig:Proper-motion-geometry})
the angle between the source velocity $\mathbf{v}_{s}$ and the direction
of arrival of a relativistic signal (photon or particle) and with
$\theta$ the corresponding angle for a non-relativistic particle
and let us assume a uniform motion. All velocities are in the observer
rest frame. In order for the two signals to arrive at approximately
the same time we require 
\begin{equation}
\frac{1}{v\sin\theta}-\frac{\cot\theta}{v_{s}}\approx\frac{1}{\sin\theta_{c}}-\frac{\cot\theta_{c}}{v_{s}}
\end{equation}
 for non-relativistic $v_{s}\equiv|\mathbf{v}_{s}|$. This condition
can be solved analytically to obtain $\theta(\theta_{c},v_{s},v)$.
It is more useful however to derive an approximate solution. A non-relativistic
particle will arrive therefore with a proper motion deviation angle
\begin{equation}
\Delta_{PM}=\theta_{c}-\theta
\end{equation}
 with respect to the relativistic signal. For small deviations $\Delta_{PM}$
we obtain 
\begin{equation}
\Delta_{PM}\approx\frac{(1-v)v_{s}\sin\theta_{c}}{v-v_{s}\cos\theta_{c}}\approx\frac{v_{s}\sin\theta_{c}}{v-v_{s}\cos\theta_{c}}
\end{equation}
 where the last step applies for $v\ll1$. Note that $v$ is always
larger that $v_{s}\cos\theta_{c}$ due to the composition of velocities,
so $\Delta_{PM}>0$. Finally, if the source is moving slowly with
respect to the particle velocity (a reasonable assumption since the\textbf{
}velocity dispersion in the Milky Way at the Sun's location is around
50-100 km/sec (see e.g. \cite{2011MNRAS.412.1203N}) we obtain 
\begin{equation}
\Delta_{PM}\approx\frac{v_{s}}{v}\frac{\xi_{PM}}{r_{s}}=\frac{\nu r_{s}}{cv}\label{eq:delta-pma}
\end{equation}
 in terms of the impact parameter $\xi_{PM}$ and the distance to
the source $r_{s}$ or in terms of the proper motion $\nu\equiv cv_{s}\sin\left(\theta_{c}\right)/r_{s}$
(the velocity of light appears explicitly now because $\nu$ is the
angular velocity). Unless the source proper motion is negligible,
the particles will arrive to the observer aligned on a direction which
is intermediate between the lensing direction (source-lens) and the
source proper motion vector. Assuming a proper motion of $\mu$arcsec$/$year
we have 
\begin{equation}
\Delta_{PM}\approx(9\deg)\mu(\frac{r_{s}}{10kpc})(\frac{v}{10^{-3}})^{-1}\label{eq:delta-pma-1}
\end{equation}
i.e. a typical deviation of ten degrees from the direction induced
by the non-relativistic lens. In general the resulting detection trail
will be a combination of the lensing and proper motion effect, it
will point in a random direction and for large $v$ it will no longer
be an exact straight line due to the different behavior with respect
to $v$. In principle however the parameters $\mu,r_{s}$ of a given
galactic source can be estimated through direct astrometric observation
and utilized to subtract the proper motion from the detection map.

Let us finally notice that the particle velocity $v$ and arrival
angle $\theta$ in the earth rest frame are different from the corresponding
quantities in the source rest frame. Suppose the source has a peculiar
velocity vector $\mathbf{v_{s}}$ with respect to us. A particle is
emitted with velocity $\mathbf{u}$ in the frame in which the source
is at rest, and received with velocity $\mathbf{v}$ in our rest frame.
All velocities are in units of $c$. The relativistic addition theorem
for velocities says that 
\begin{equation}
\mathbf{v}=\frac{\mathbf{u}_{\parallel}-\mathbf{v}_{s}}{1-vu}+\frac{\mathbf{u}_{\perp}}{\gamma_{s}(1-vu)}
\end{equation}
 where $\gamma_{s}=(1-v_{s}^{2})^{-1/2}$ and 
\begin{eqnarray}
\mathbf{u}_{\parallel} & = & \mathbf{v}_{s}(\mathbf{v}_{s}\cdot\mathbf{u})/V^{2}\\
\mathbf{u}_{\perp} & = & \mathbf{u}-\mathbf{u}_{\parallel}
\end{eqnarray}
 Putting 
\begin{eqnarray}
\mathbf{u} & = & u\{\cos\theta',\sin\theta'\}\\
\mathbf{v} & = & v\{\cos\theta,\sin\theta\}
\end{eqnarray}
 we find 
\begin{eqnarray}
\sin\theta & = & \frac{u\sin\theta'}{[\gamma_{s}^{2}(v_{s}+u\cos\theta')^{2}+u^{2}\sin^{2}\theta')]^{1/2}}\\
v & = & \frac{[\gamma_{s}^{2}(v_{s}+u\cos\theta')^{2}+u^{2}\sin^{2}\theta')]^{1/2}}{\gamma_{s}(1+uv_{s}\cos\theta')}
\end{eqnarray}
 The equation for $\theta'$ can also be written as 
\begin{equation}
\tan\frac{\theta}{2}=\frac{u\sin\theta'}{\gamma_{s}(v_{s}+u\cos\theta')+[\gamma_{s}^{2}(v_{s}+u\cos\theta')^{2}+u^{2}\sin^{2}\theta')]^{1/2}}
\end{equation}
 which for $u\to1$ becomes the standard aberration equation 
\begin{eqnarray}
\tan\frac{\theta}{2} & = & \tan\frac{\theta'}{2}\left(\frac{1-v_{s}}{1+v_{s}}\right)^{1/2}
\end{eqnarray}
 (and at the same time $u=v$).

\bibliographystyle{amsplain} \bibliographystyle{amsplain} 
\end{document}